\documentclass[12pt]{iopart}
\usepackage{graphicx}
\begin{document}
\title{The fine structure of Gowdy spacetimes}
\author{David Garfinkle}
\address{Dept. of Physics, University of Guelph, Guelph, Ontario,
N1G 2W1, Canada}
\ead{david@physics.uoguelph.ca}
\begin{abstract}
The approach to the singularity in Gowdy spacetimes consists of 
velocity term dominated behavior, except at a set of isolated points.
At and near these points, spiky features grow.  This paper reviews 
what is known about these spikes.

\end{abstract}
\maketitle
\section{Introduction}

There have been several investigations, both analytical and numerical,
of the approach to the singularity in inhomogeneous cosmologies.\cite{
vinceandjim}-\cite{ringstrom}
The
most extensively studied class of these spacetimes is the class of 
Gowdy spacetimes\cite{gowdy} on ${T^3} \times R$.  Theorems due to 
Isenberg and Moncrief\cite{vinceandjim} showed that the polarized 
Gowdy spacetimes are asymptotically velocity term dominated (AVTD).
It was then natural to ask whether this is also true of the more
general unpolarized Gowdy spacetimes.  This was a more difficult question
since the relevant equations are linear in the polarized case but
nonlinear in the unpolarized case.  To address this question, Berger
and Moncrief\cite{beverlyandvince} performed numerical simulations
of the approach to the singularity in Gowdy spacetimes.  They obtained
the following curious result: the behavior is AVTD except at a set of 
isolated points.  At and near these exceptional points, features appear
that grow ever steeper and narrower as the singularity is approached.  
Berger and Moncrief referred to these features as ``spiky features,''
though for brevity we will simply call them ``spikes.''  These spikes
form a ``fine structure'' on the simpler and more prevalent behavior
of the rest of the spacetime.

The purpose of this paper is to review what is known about these spikes.
Section 2 presents the basic equations governing the Gowdy spacetimes
on ${T^3} \times R$.  The next three sections concern a variety of methods
and results, with numerical methods and results presented in section
3, analytical approximations in section 4 and mathematical results in
section 5.  Conclusions are given in section 6. 

\section{Equations}

The Gowdy metric on ${T^3}\times R$ takes the form
\begin{equation}
d {s^2} = {e^{\lambda /2}}{t^{-1/2}}(-d{t^2}+d{x^2})+t[{e^P}
{{(dy+Qdz)}^2}+{e^{-P}}d{z^2}]
\label{gowdymetric}
\end{equation}
where $P, Q$ and $\lambda $ are functions of $t$ and $x$.  
The $T^3$ spatial topology is imposed by having 
$0\le x,y,z \le 2 \pi$ and having $P, Q$ and $\lambda$ be periodic
functions of $x$.
The vacuum
Einstein field equations split into ``evolution'' equations for
$P$ and $Q$ 
\begin{eqnarray}
{P_{,tt}}+{t^{-1}}{P_{,t}}-{P_{,xx}}+{e^{2P}}({Q_{,x} ^2}-{Q_{,t} ^2}
)=0
\label{evolvePt}  \\
{Q_{,tt}}+{t^{-1}}{Q_{,t}}-{Q_{,xx}}+2({P_{,t}}{Q_{,t}}-{P_{,x}}{Q_{,x}}
)=0 
\label{evolveQt}
\end{eqnarray}
and ``constraint'' equations for $\lambda$
\begin{eqnarray}
{\lambda _{,t}}=t [ {P_{,t} ^2} +{P_{,x} ^2} +{e^{2P}} ( {Q_{,t} ^2}
+ {Q_{,x} ^2})] 
\label{lambdat} \\
{\lambda _{,x}}=2 t ( {P_{,x}}{P_{,t}}+{e^{2P}}{Q_{,x}}{Q_{,t}})
\label{lambdax}
\end{eqnarray}
(here ${_{,a}}=\partial /\partial a$).
The constraint equations determine $\lambda$ once $P$ and $Q$ are known.
The integrability conditions for the constraint equations are satisfied 
as a consequence of the evolution equations.  Since the evolution equations
do not depend on $\lambda$ there is essentially a complete decoupling 
of constraints from evolution equations.  Therefore, for the purposes of this 
paper we will treat only equations (\ref{evolvePt}-\ref{evolveQt}).  
The only restriction that the constraints place on initial data for
equations (\ref{evolvePt}-\ref{evolveQt}) is the following: since
$\lambda $ at $x=0$ is the same as $\lambda $ at $x=2 \pi$, it follows that
the integral from $0$ to $2\pi$ of the right hand side of equation  
(\ref{lambdax}) must vanish.  We require that this restriction is satisfied by
the initial data for equations (\ref{evolvePt}-\ref{evolveQt}) and then
these equations insure that the restriction is also satisfied at 
subsequent times.

The singularity is at $t=0$.  It is often helpful to introduce the
coordinate $\tau \equiv - \ln t$.  Thus the singularity is approached
as $\tau \to \infty$.  In terms of this coordinate, the evolution
equations (\ref{evolvePt}-\ref{evolveQt}) become
\begin{eqnarray}
{P_{,\tau \tau}}-{e^{2P}}{Q_{,\tau} ^2}-{e^{-2\tau}}{P_{,xx}}+
{e^{2(P-\tau )}}{Q_{,x} ^2} = 0
\label{evolvePtau} \\
{Q_{,\tau \tau}}+2{P_{,\tau}}{Q_{,\tau}}-{e^{-2\tau}}({Q_{,xx}}+
2{P_{,x}}{Q_{,x}})=0
\label{evolveQtau}
\end{eqnarray}

Equations (\ref{evolvePtau}-\ref{evolveQtau}) are the equations of motion 
corresponding to the Hamiltonian $H={H_K}+{H_V}$ where
\begin{eqnarray}
{H_K}= {1 \over 2} {\int _0 ^{2 \pi}} d x ( {\pi _P ^2} +
{e^{-2P}}{\pi _Q ^2}),
\label{Kinetic} \\
{H_V} = {1 \over 2} {e^{-2\tau}} {\int _0 ^{2 \pi}} d x 
({P_{,x} ^2} + {e^{2 P}} {Q_{,x} ^2}).
\label{Potential}
\end{eqnarray} 

One can also put the evolution equations (\ref{evolvePt}-\ref{evolveQt})
in characteristic form: define null coordinates $\xi \equiv t+x$ and
$\eta \equiv t-x$ and characteristic variables $A, B, C$ and $D$ given by
\begin{eqnarray}
A = (\xi + \eta ){P_{,\eta}},
\label{Adef} \\
B = (\xi + \eta ){P_{,\xi}},
\label{Bdef} \\
C = (\xi + \eta ){e^P}{Q_{,\eta}},
\label{Cdef} \\
D = (\xi + \eta ){e^P}{Q_{,\xi}}.
\end{eqnarray}
Then equations (\ref{evolvePt}-\ref{evolveQt}) become
\begin{eqnarray}
{A_{,\xi}}={{(\xi + \eta )}^{-1}}[CD+(A-B)/2]
\label{Au} \\
{B_{,\eta}} = {{(\xi + \eta )}^{-1}}[CD+(B-A)/2]
\label{Bw} \\
{C_{,\xi}}={{(\xi + \eta )}^{-1}}[-AD+(C-D)/2]
\label{Cu} \\
{D_{,\eta}}={{(\xi + \eta )}^{-1}}[-BC+(D-C)/2]
\end{eqnarray}

As we will see in the next section, each of these different ways of 
writing the evolution equations has its uses for numerical simulations.

\section{Numerical methods and results}

We now consider numerical simulations of the evolution equations
for $P$ and $Q$.  Since we wish to examine the behavior as the
singularity is approached, it is most convenient to simulate
equations (\ref{evolvePtau}-\ref{evolveQtau}) and examine the
behavior for large $\tau$.  One simple numerical method is to use
centered differences for spatial derivatives and the iterated
Crank-Nicholson (ICN) method\cite{icn} 
for time evolution.  (Spikes were 
first found\cite{beverlyandvince} using the symplectic method
to be described below.  However, the ICN method has been 
used\cite{S2} to simulate Gowdy spacetimes with spatial topology
${S^2}\times {S^1}$). 
The ICN method is in the class of so called ``predictor-corrector''
methods: first taking a crude approximation to the evolution and 
then refining it.  The symplectic method is in the class of 
``operator splitting'' methods: splitting the equations of motion 
into two pieces, each of which is evolved separately.
Here centered 
differences means that for a quantity $F$ at spatial grid point
$i$ we have ${F_{,x}}\to ({F_{i+1}}-{F_{i-1}})/(2\Delta x)$ and
${F_{,xx}}\to ({F_{i+1}+{F_{i-1}}-2{F_i})/{{(\Delta x})}^2}$.
The ICN method is the following: we put the equations in the form
${S_{,\tau}}=W$ ({\it i.e.} first order in time) by making 
$P_{,\tau}$ and $Q_{,\tau}$ extra variables.  Then we evolve the
variables $S$ from time step $n$ to time step $n+1$ using
\begin{equation}
{S^{n+1}}={S^n}+ {{\Delta \tau}\over 2} [ W({S^n})+W({S^{n+1}})]
\label{icn}
\end{equation}
We choose $S^n$ as our initial guess for $S^{n+1}$ and use this 
guess in the right hand side of equation (\ref{icn}) to produce 
a better guess.  This process is iterated 3 times to evolve to the
next time step.     

\begin{figure}
\includegraphics{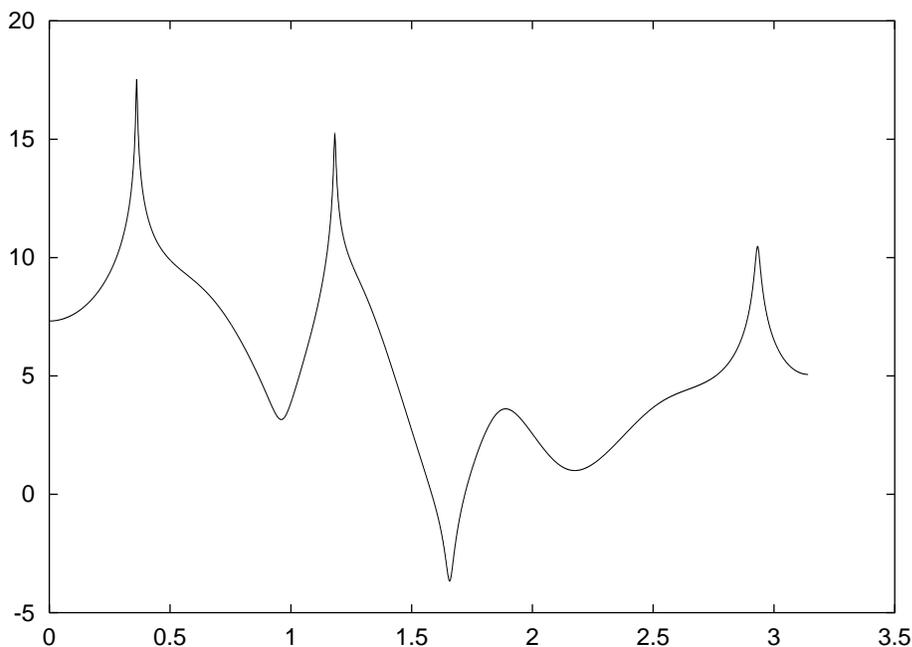}
\caption{\label{fig1}Plot of $P$ {\it vs} $x$ for ${v_0}=5$ at
$\tau =10$. Here $0 \le x \le \pi$}
\end{figure}

\begin{figure}
\includegraphics{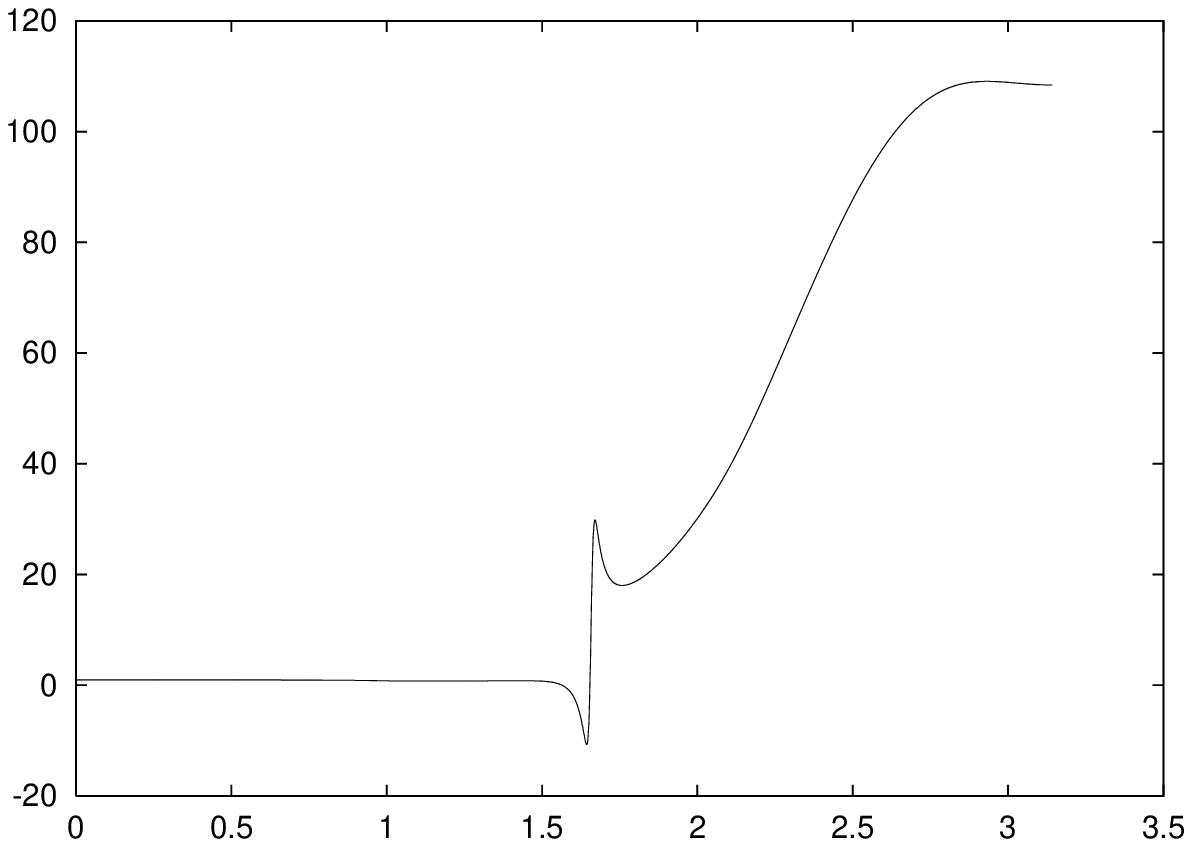}
\caption{\label{fig2}Plot of $Q$ {\it vs} $x$ for ${v_0}=5$ at
$\tau =10$.  Here $0 \le x \le \pi$}
\end{figure}

We wish to consider small scale structure generated by the approach
to the singularity.  For that reason we choose initial data that
itself has as little small scale structure as possible.  Following
references\cite{beverlyandvince,beverlyandme} we use the data
$P=0, \, {P_{,\tau}}={v_0}\cos x, \, Q = \cos x, {Q_{,\tau}}=0$
at time $\tau =0$.  (Here $v_0$ is a constant).  As argued in 
reference\cite{beverlyandme} the type of fine structure generated by this    
family of initial data should be generic.  Figures (\ref{fig1})
and (\ref{fig2}) show the
evolution to $\tau=10$ of these data for ${v_0}=5$.

\begin{figure}
\includegraphics{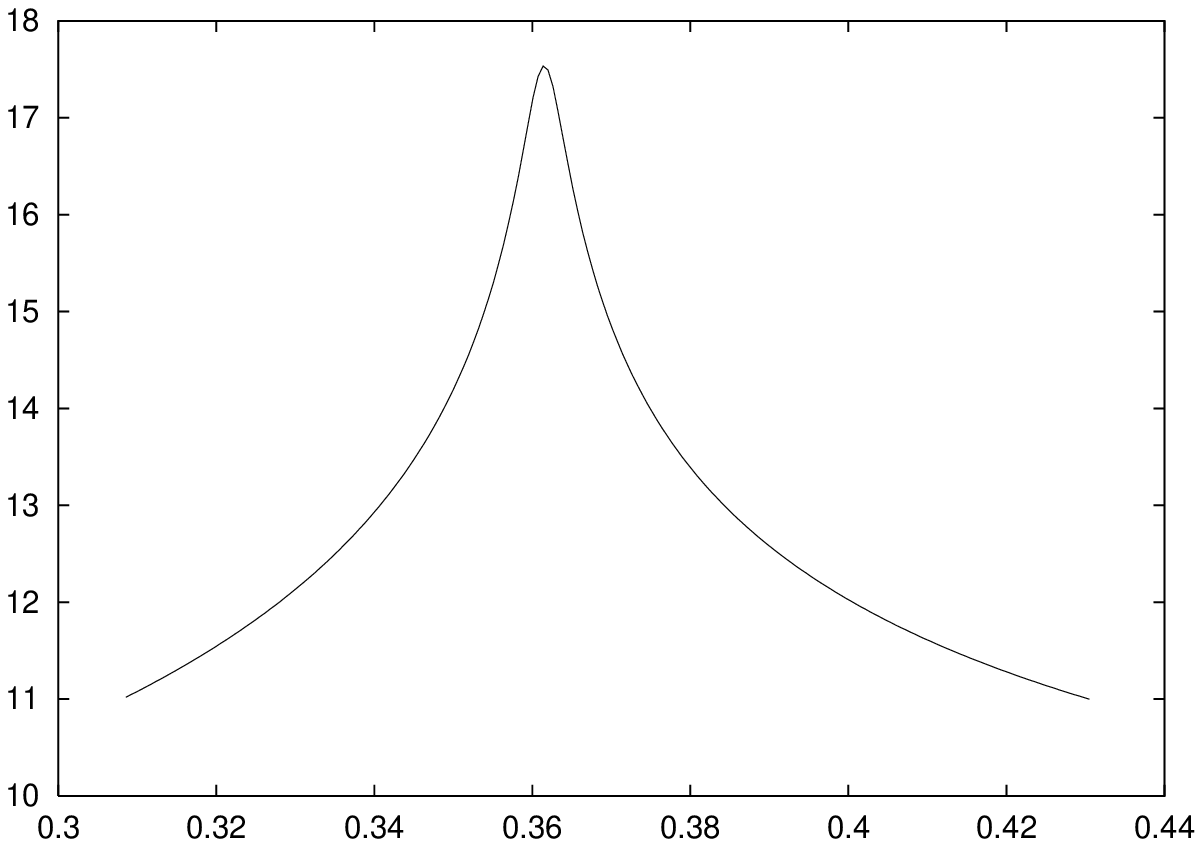}
\caption{\label{fig3}Plot of $P$ {\it vs} $x$ for ${v_0}=5$ at
$\tau =10$. This is a closer look at the 
true spike at $x\approx 0.36$}
\end{figure}

\begin{figure}
\includegraphics{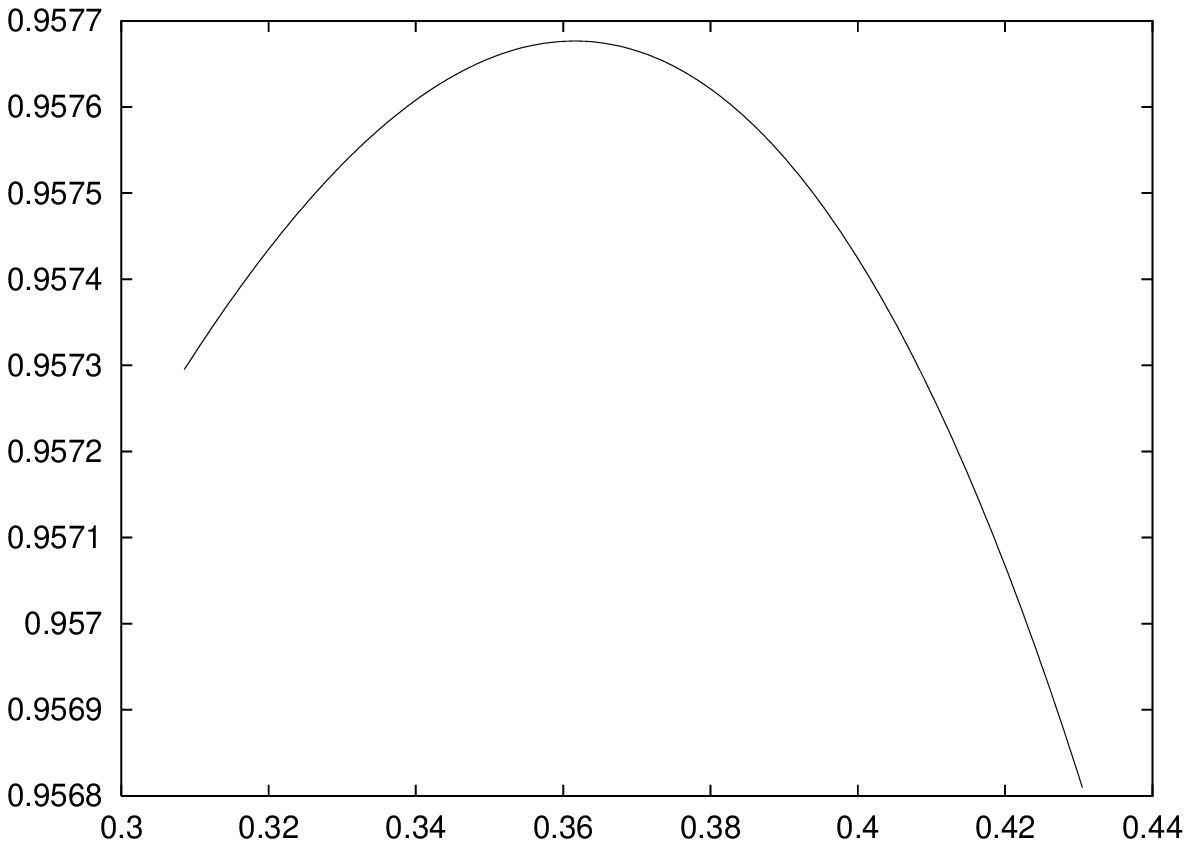}
\caption{\label{fig4}Plot of $Q$ {\it vs} $x$ for ${v_0}=5$ at
$\tau =10$.  This is a closer look at the true spike at 
$x \approx 0.36$}
\end{figure}

\begin{figure}
\includegraphics{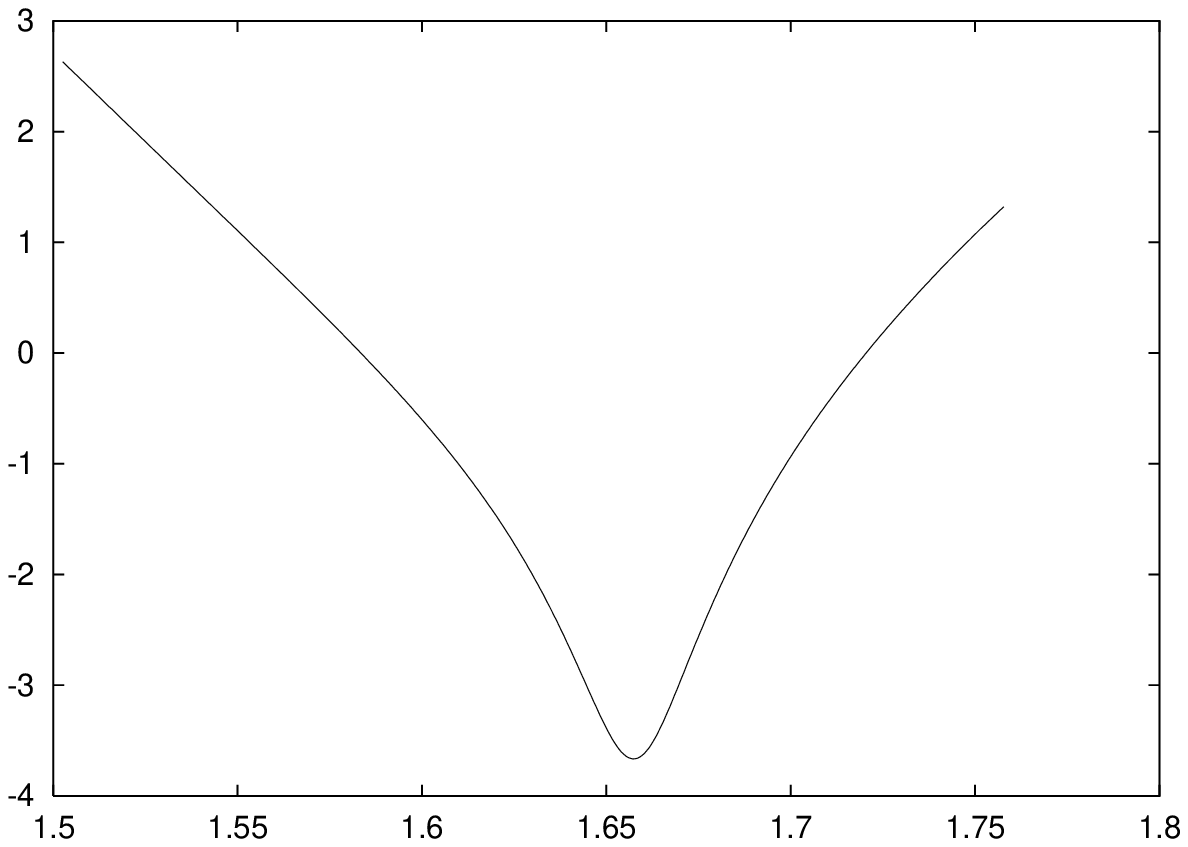}
\caption{\label{fig5}Plot of $P$ {\it vs} $x$ for ${v_0}=5$ at
$\tau =10$. This is a closer look at the
false spike}
\end{figure}

\begin{figure}
\includegraphics{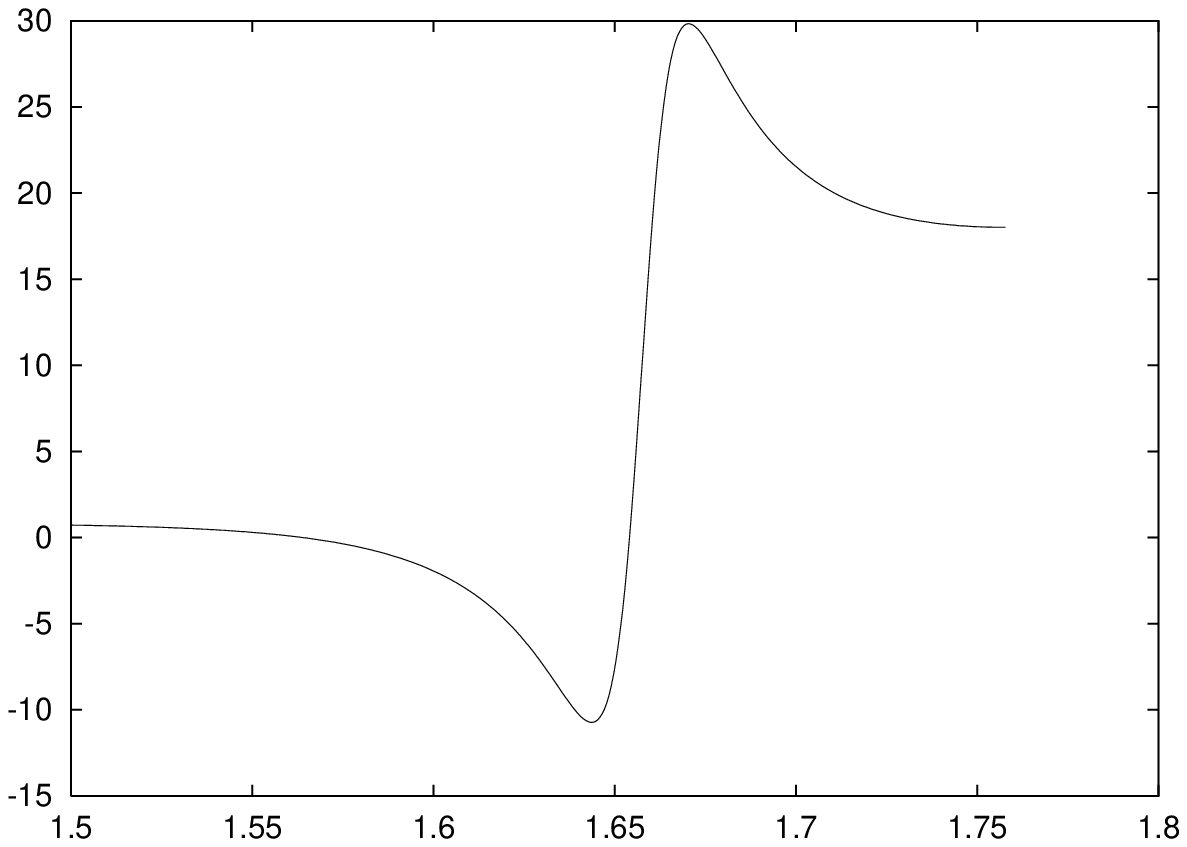}
\caption{\label{fig6}Plot of $Q$ {\it vs} $x$ for ${v_0}=5$ at
$\tau =10$.  This is a closer look at the false spike}
\end{figure}

\begin{figure}
\includegraphics{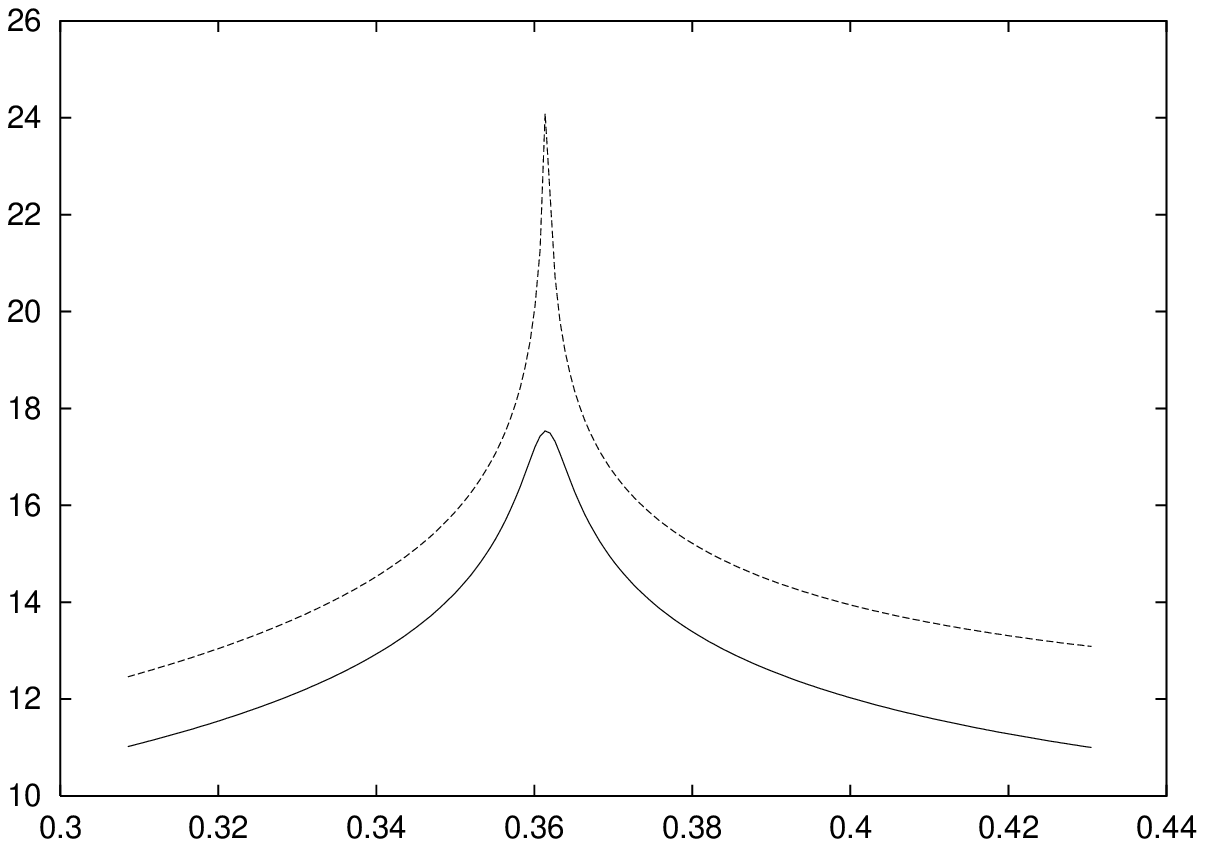}
\caption{\label{fig7}Plot of $P$ {\it vs} $x$ for ${v_0}=5$ at
$\tau =10$ (solid line) and $\tau =15$ (dashed line). This is a 
close look at the
true spike at two different times.  Note that the spike
is higher and more narrow at the later time.}
\end{figure}

\begin{figure}
\includegraphics{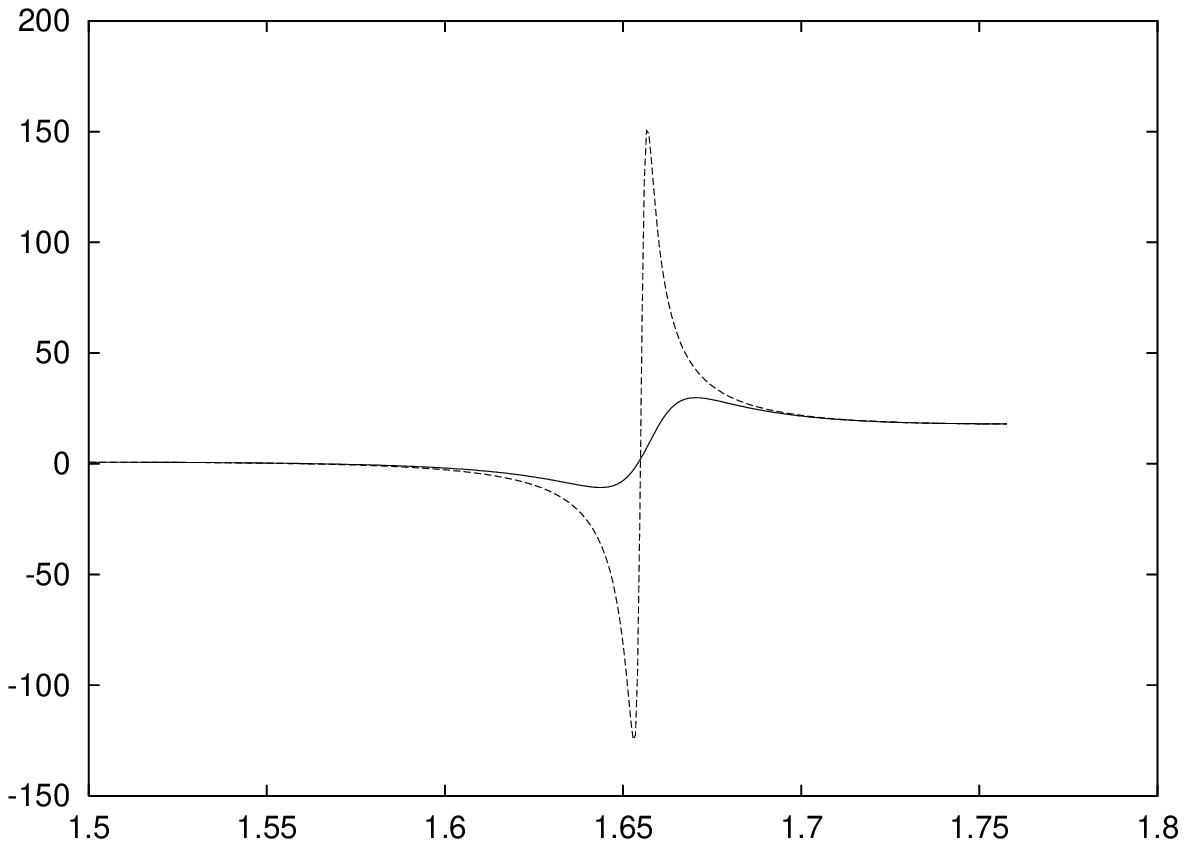}
\caption{\label{fig8}Plot of $Q$ {\it vs} $x$ for ${v_0}=5$ at
$\tau =10$ (solid line) and $\tau =15$ (dashed line). This is a 
close look at the
false spike at two different times.  Note that the spike
is higher and more narrow at the later time.}
\end{figure}

From these plots it is apparent that there are narrow upward pointing 
features in $P$ at $x \approx 0.36, \, x \approx 1.2 $ and $x \approx
2.9$.  In addition, there is a narrow downward pointing feature
in $P$ at $x \approx 1.66$ which coincides with the location of a narrow
feature in $Q$.  For reasons that will become apparent in section 5,
we adopt the notation of reference\cite{alanandmarsha} and call the
upward pointing features in $P$ ``true spikes'' and the downward pointing
features in $P$ along with narrow feature in $Q$ a ``false spike.''

Figures (\ref{fig3}) and (\ref{fig4}) provide a closer 
look at the true spike at $x\approx 0.36$
while figures (\ref{fig5}) and (\ref{fig6}) do the same 
for the false spike.  Note that
$Q$ has an extremum at the position of the true spike.

The spikes become steeper and narrower as the singularity is approached.
Figure (7) shows $P$ at the true spike at $x \approx 0.36$ at 
$\tau =10$ and $\tau =15$.  Figure (\ref{fig8}) 
shows $Q$ at the false spike
at $\tau =10$ and $\tau =15$.   

The ICN method that we have described is second order accurate.  
However, one might want higher order accuracy, and this is provided
by the symplectic methods used in\cite{beverlyandvince}.  The
symplectic method\cite{suzuki} works as follows: begin with a
system whose equations of motion come from a Hamiltonian of the
form $H={H_K}+{H_V}$ where the sub-hamiltonians $H_K$ and
$H_V$ each correspond to a system that can be solved in closed
form.  (Note that this condition is satisfied by the Gowdy evolution
equations (\ref{evolvePtau}-\ref{evolveQtau}) where the
corresponding sub-hamiltonians are given in equations
(\ref{Kinetic}-\ref{Potential})).  Let ${{\cal U}_K}(\Delta \tau )$
be the evolution operator that corresponds to evolving the system
with Hamiltonian $H_K$ for a time $\Delta \tau$.  Correspondingly 
define  ${{\cal U}_V}(\Delta \tau )$.  Now consider the operator
${{\cal U}_{(2)}}(\Delta \tau )$ defined by
\begin{equation}
{{\cal U}_{(2)}}(\Delta \tau ) = {{\cal U}_K}(\Delta \tau /2)
{{\cal U}_V}(\Delta \tau ){{\cal U}_K}(\Delta \tau /2).
\label{U2def}
\end{equation}         
In words, ${{\cal U}_{(2)}}(\Delta \tau )$ evolves using $H_K$ for half
a time step, then using $H_V$ for a full time step and then using 
$H_K$ for half a time step.  One can show\cite{suzuki} that ${\cal U}_{(2)}$
is a second order accurate operator for the full system described by
$H={H_K}+{H_V}$.  Furthermore, using ${\cal U}_{(2)}$ one can construct
evolution operators of any accuracy one wishes.  The 
fourth order accurate operator ${\cal U}_{(4)}$ is given by
\begin{equation}
{{\cal U}_{(4)}}(\Delta \tau ) = {{\cal U}_{(2)}}(s \Delta \tau )
{{\cal U}_{(2)}}[(1-2s)\Delta \tau ] {{\cal U}_{(2)}}(s\Delta \tau )
\label{U4def}
\end{equation}
where $s=1/(2-{2^{1/3}})$.  The general nth order accurate operator 
${\cal U}_{(n)}$ is built recursively from lower order ${\cal U}_{(n)}$
in similar ways. 

To take advantage of nth order accuracy in time, the codes must also
have nth order accuracy in space.  Thus the second order accurate
centered differences must be replaced by more complicated finite
differences that are nth order accurate.  There is another advantage
besides higher order accuracy in using the symplectic method to study
the approach to the singularity: as the singularity is approached, 
$H_V$ appears to become small compared to $H_K$.  If $H_V$ were completely
negligible, then the operator ${\cal U}_{(2)}$ (and therefore
${\cal U}_{(4)}$ and all higher order ${\cal U}_{(n)}$)      
would be exact.  Thus as the singularity is approached, the symplectic
method may be even more accurate than one might guess from the 
formal order of accuracy.  (This fact has been used in simulations of
Mixmaster spacetimes\cite{bgs}. )

A numerical method will be accurate only if there are enough spatial
points to resolve all features.  However, the spikes grow ever
narrower; thus for any fixed resolution there will come a time
when that resolution is not sufficient to resolve the spikes.
Indeed the simulations of \cite{beverlyandvince,beverlyandme}
are often run so long that some spikes are not resolved; however
it is argued in\cite{beverlyandme} that this does not affect the
accuracy of the simulation in regions away from the spikes.

To resolve the spikes for a long time, it is better to use a grid
that does not have a fixed resolution.  One way to do this is
to use the technique of adaptive mesh refinement 
(AMR).\cite{bergerandoliger}
This technique adds extra mesh points in places where the existing
mesh is not sufficiently fine to resolve the features.  AMR was applied
to the Gowdy spacetimes by Hern and Stewart.\cite{hs}  Unfortunately,
AMR codes are quite involved and difficult to write.  
Moreover, the full machinery of AMR may not be needed
since it appears that 
in a given Gowdy spacetime there is a
limited number of spikes and that their positions change very little
as the singularity is approached.  One can then construct a mesh
that is finer in the places where one knows the spikes 
will be.\cite{marshaphd}

Alternatively, one can study a single spike using a characteristic
method.\cite{meandmarsha}  Here, the initial data surface is an
outgoing light ray and one evolves along ingoing light rays.  One can 
choose the last grid point to correspond to the light ray
that hits the center of the spike at the singularity.  As a consequence
of this choice, the grid shrinks as the evolution proceeds in such
a way that the grid is always of an appropriate size to resolve the
spike.  

\section{Analytical approximations}

We now turn to analytical approximations to the equations of motion
(\ref{evolvePtau}-\ref{evolveQtau}).  Our treatment closely follows 
that of\cite{beverlyandme}.  One obvious approximation
is the following: since the approach to the singularity is
$\tau \to \infty$, it is natural to neglect terms in equations
(\ref{evolvePtau}-\ref{evolveQtau}) proportional to 
$e^{-2\tau}$.  Dropping such terms results in the following
equations:
\begin{eqnarray}
{P_{,\tau \tau}} - {e^{2P}}{Q_{,\tau} ^2} = 0 ,
\label{VTDP} \\
{Q_{,\tau \tau}} + 2 {P_{,\tau}}{Q_{,\tau}}=0
\label{VTDQ}
\end{eqnarray}
These equations are called the velocity term dominated (VTD) equations.
(Note that the VTD equations are the equations of motion corresponding
to the sub-hamiltonian $H_K$).  A solution of equations
(\ref{evolvePtau}-\ref{evolveQtau}) is called asymptotically velocity
term dominated (AVTD) if there is a solution of the VTD equations that it 
approaches as $\tau \to \infty$.

\begin{figure}
\includegraphics{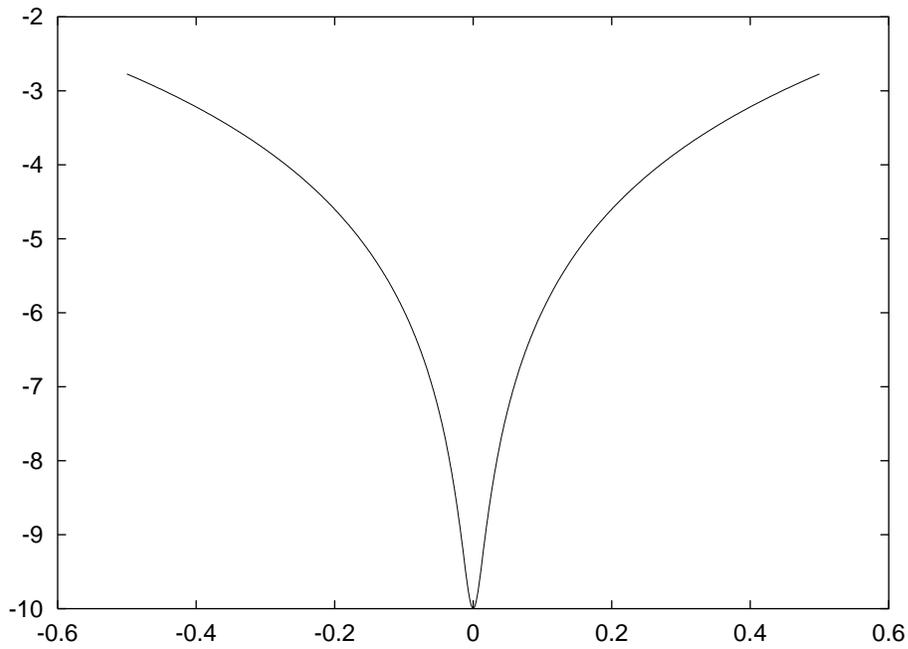}
\caption{\label{fig9}Plot of $P$ {\it vs} $x$ for
the analytic approximation, equation (\ref{falseP}) for a false spike}
\end{figure}

\begin{figure}
\includegraphics{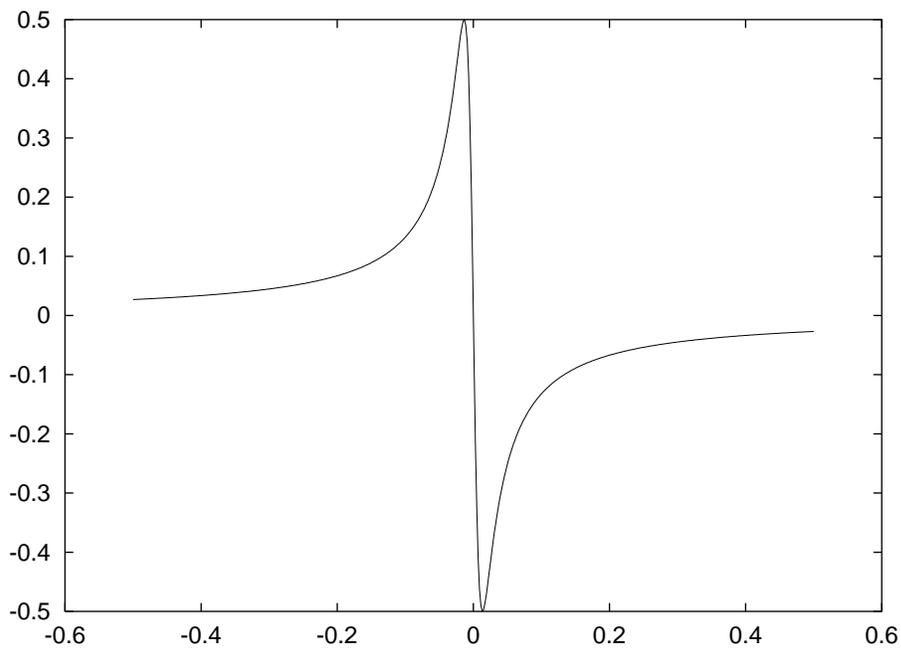}
\caption{\label{fig10}Plot of $Q$ {\it vs} $x$ for
the analytic approximation, equation (\ref{falseQ}) for a false spike}
\end{figure}

The VTD equations can be solved in closed form with the general solution
given by
\begin{eqnarray}
P=p+\ln [\cosh v \tau + \cos \psi \sinh v \tau ],
\label{VTDPsoln} \\
Q = q + {{{e^{-p}}\sin \psi \tanh v \tau}\over 
{1 + \cos \psi \tanh v \tau}}.
\label{VTDQsoln}
\end{eqnarray}
Here $v \ge 0$ and the quantities $p, \, q, \, v$ and $\psi$ 
are functions of $x$.  The behavior of the false spikes follows from
equations (\ref{VTDPsoln}-\ref{VTDQsoln}).  For $\cos \psi \ne -1$ the
behavior of these solutions as $\tau \to \infty $ is $P \to {\bar p}
+ v \tau $ and $Q \to {\bar q}$ where $\bar p$ and $\bar q$ are 
functions of $x$ given by 
${\bar p} = p + \ln [(1+\cos \psi )/2]$ and
${\bar q} = q + {e^{-p}} \sin \psi /(1+\cos \psi )$.
Let $x_1$ be a point where $\cos \psi = - 1$.  Then clearly there is some
steep behavior near $x_1$.  Let a subscript $1$ denote the value of the function at $x_1$ 
and let ${\psi _1}'$ denote $\psi _{,x}$ at $x_1$.  Then near $x_1$ 
and for large $\tau $ it follows from equations (\ref{VTDPsoln}-
\ref{VTDQsoln}) that 
\begin{eqnarray}
P \approx {p_1} - {v_1} \tau + \ln [ 1+ {{({{\psi _1}'})}^2}
{e^{2 {v_1}\tau}}{{(x-{x_1})}^2}/4]
\label{falseP} \\ 
Q \approx {q_1} - {{{e^{-{p_1}}}{{\psi _1}'}(x-{x_1})}\over
{2 {e^{- 2{v_1}\tau}}+{{({{\psi _1}'})}^2}{{(x-{x_1})}^2}/2}}
\label{falseQ}
\end{eqnarray}
Equations (\ref{falseP}-\ref{falseQ}) provide an analytic approximation
to the behavior of false spikes.  
Figures (\ref{fig9}-\ref{fig10}) show the behavior of 
$P$ and $Q$ respectively for these formulas.  Here ${p_1}=5,\, 
{q_1}=0,\, {v_1}=0.5,\, {x_1}=0,\, {{\psi _1}'}=1 $ and $\tau=10$.  

\begin{figure}
\includegraphics{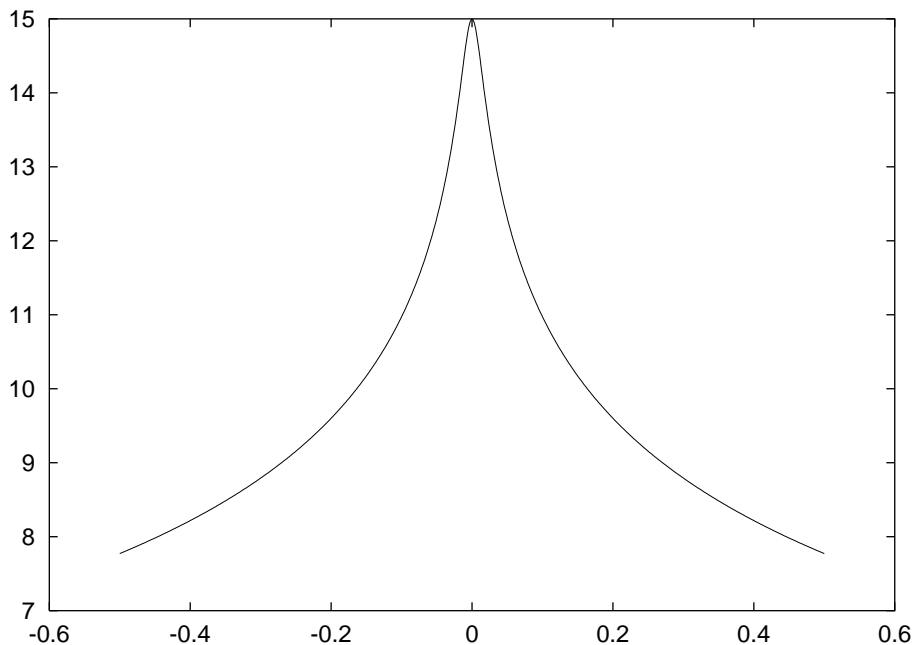}
\caption{\label{fig11}Plot of $P$ {\it vs} $x$ for 
the analytic approximation, equation (\ref{trueP}) for a true spike}
\end{figure}

To treat the behavior of the true spikes, we first consider the
validity of the VTD approximation by using the ``method of 
consistent potentials'' (MCP).\cite{boroandvince} That is,
we assume the large $\tau $ behavior $P\to {\bar p} + v \tau , \,
Q \to {\bar q}$ and see whether the terms in equations (\ref{evolvePtau}
-\ref{evolveQtau}) that we have neglected are in fact negligible.  
The term ${e^{2(P-\tau )}}{Q_{,x} ^2}$ has behavior
${e^{2[{\bar p}+(v-1)\tau ]}}{{\bar q}_{,x} ^2}$.
Thus this term is negligible only if $v < 1$.  The numerical simulations
of\cite{beverlyandvince,beverlyandme} show that in fact $v$ 
evolves to a value less
than 1 except at the true spikes.  To find an explanation for this 
behavior we must consider what happens when $v$ is initially greater
than 1.  We are still justified in using equation (\ref{VTDQ}) which
yields ${Q_{,\tau}}= {\pi _Q}{e^{-2P}}$ where $\pi _Q$ is a function
of $x$.  This turns the term $- {e^{2P}}{Q_{,\tau} ^2}$ in equation
(\ref{evolvePtau}) into $- {e^{-2P}}{\pi _Q ^2}$ which can be neglected
since $P$ is growing.  We can also approximate $Q(\tau ,x)$ by
${\bar q}(x)$.  Thus we are led to approximate equation 
(\ref{evolvePtau}) by
\begin{equation}
{P_{,\tau \tau}} + {e^{2(P-\tau )}}{{\bar q}_{,x} ^2} = 0.
\label{Phighv}
\end{equation}
This equation can be solved in closed form to yield
\begin{equation}
P = p + \tau - \ln [ \cosh w \tau - \cos \phi \sinh w \tau ] .
\label{Phighvsoln}
\end{equation}
Here $w>0$ and $p,\, w$ and $\phi$ are functions of $x$ and we have
${{\bar q}_{,x}}={e^{-p}}w\sin \phi$.  For $\cos \phi \ne 1 $ the
large $\tau$ behavior of this solution is $P \to {\bar p} + v \tau$
where ${\bar p} = p - \ln [ (1-\cos \phi )/2]$ and $v=1-w$.
Let $x_2$ be a point where $\cos \phi = 1$ and use a subscript 2 to
denote the value of the function at $x_2$.  Then near $x_2$ and for large $\tau$ if follows 
from equation (\ref{Phighvsoln}) that 
\begin{equation}
P \approx {p_2} + (1+{w_2})\tau - \ln [ 1 + {{({{\phi _2}'})}^2}
{e^{2 {w_2}\tau}}{{(x-{x_2})}^2}/4]
\label{trueP}
\end{equation}
Equation (\ref{trueP}) provides an analytic approximation to the
behavior of true spikes.  Figure (\ref{fig11}) shows the behavior of $P$ for
this formula.  Here ${p_2} =0,\, {w_2}=0.5,\, {x_2}=0,\, 
{{\phi _2}'}=1$ and $\tau = 10$.

The dynamics of spike formation can be understood as follows: Define 
the ``potentials'' ${V_1} = {\pi _Q ^2} {e^{-2P}}$ and
${V_2} = {e^{2(P-\tau )}}{Q_{,x} ^2}$.  Then the behavior of 
$P$ can be regarded as the the dynamics of a particle in these
potentials.  If ${P_{,\tau}}<0$ then a ``bounce'' off of $V_1$ 
will give rise to a transition ${P_{,\tau }}\to - {P_{,\tau}}$
which will make $P_{,\tau}$ positive.  If ${P_{,\tau }} > 1$ then
a bounce off of $V_2$ will give rise to a transition ${P_{,\tau}}
\to 1-{P_{,\tau}}$ and thus ${P_{,\tau}} < 1$.  Eventually, 
after a finite number of bounces $P_{,\tau}$ will be in the range
$0<{P_{,\tau}}<1$.  This occurs provided that both potentials
$V_1$ and $V_2$ are nonvanishing.  At a point $x_1$ where ${V_1}=0$
a false spike can form since ${P_{,\tau}}<0$ is allowed.  At a 
point $x_2$ where ${V_2}=0$ a true spike can form since ${P_{,\tau}}
>1$ is allowed.  

We now consider the validity of the approximation (\ref{trueP}).
Using the MCP we find that the term ${e^{-2\tau}}{P_{,xx}}$ goes as
$e^{2(w-1)\tau}$.  This term is thus negligible provided that
$w<1$ or equivalently that $P_{,\tau}$ at the center of the spike
is less than 2.  This leaves open the question of what happens to 
``high velocity'' spikes, {\it i.e.} those for which $P_{,\tau}$ 
at the center of the spike is initially larger than 2.  This
question was answered recently in the numerical work 
of\cite{meandmarsha}.  The simulations of \cite{meandmarsha}
show that initially high velocity spikes are forced by the term
${e^{-2\tau}}{P_{,xx}}$ to lower velocity.  In the end, either
the spike itself disappears or the velocity at the center of
the spike is brought to the range $1<{P_{,\tau}}<2$.  

\section{Mathematical results}

Here we consider two mathematical questions about spikes: (1)
are spikes geometrical features or coordinate artifacts? and
(2) to what extent are solutions with spikes rigorously proven
to exist?  This section is essentially a summary of the results
of\cite{alanandmarsha}.  The reader should see that work and
references therein for a more detailed treatment.  

So far we have presented the spikes in terms of the behavior of
$P$ and $Q$ as functions of $\tau$ and $x$.  That is, we have
considered the coordinate dependence of metric components.  One
might then wonder whether the spikes are simply artifacts of 
the coordinate system in which the metric (\ref{gowdymetric}) 
is expressed.  However, the Gowdy coordinates and metric components
themselves have geometric meaning in terms of the symmetries
of the spacetime.  The coordinate $t$ is defined by the area of
each symmetry $T^2$ being $4{\pi ^2} t$.  The coordinate $x$ 
is the harmonic function conjugate to $t$ (see Chapter 3, problem 2
of \cite{wald}).  The vector fields ${(\partial /\partial y)}^a$    
and ${(\partial /\partial z)}^a$ are Killing fields.  Furthermore,
$P$ and $Q$ can be expressed in terms of inner products of Killing
fields.  Nonetheless, there is a restricted coordinate freedom that
preserves this relation between coordinates and geometry.  In particular,
consider the transformation that simply switches the coordinates
$y$ and $z$.  This leaves the metric in the form of (\ref{gowdymetric})
but changes the pair $(P,Q)$ to $({\tilde P},{\tilde Q})$ given by
\begin{eqnarray}
{e^{-{\tilde P}}}={{e^{-P}}\over{{Q^2}+{e^{-2P}}}},
\label{Ptilde} \\
{\tilde Q} = {Q \over{{Q^2}+{e^{-2P}}}}.
\label{Qtilde}
\end{eqnarray}
Following \cite{alanandmarsha} we will use the notation
$({\tilde P},{\tilde Q})={\rm I}(P,Q)$ where the ``I'' is for inversion.
Now suppose that $(P,Q)$ is a solution of the VTD equations without
spikes and which therefore has the large $\tau$ behavior $P \to {\bar p}
+ v \tau $ and $Q \to {\bar q}$.  Then $\tilde Q$ has the large $\tau$
behavior
\begin{equation}
{\tilde Q} \to {{\bar q} \over {{{\bar q}^2}+{e^{- 2 {\bar p}}}{e^{
- 2 v \tau}}}}.
\label{IQfalse} 
\end{equation}
Therefore $\tilde Q$ has a false spike at points where $\bar q$ vanishes.
Thus false spikes are not geometric quantities since they can be made
to appear and disappear by a coordinate transformation.  

In contrast true spikes can be shown to be true geometric objects.  
This is done by examining the behavior of curvature invariants at the
spikes as the singularity is approached.  The asymptotic behavior of 
curvature invariants as the singularity is approached is different at
a true spike than at nearby points.

We now turn to a different type of transformation that takes a
solution $(P,Q)$ of equations (\ref{evolvePtau}-\ref{evolveQtau})
and produces a solution $({\bar P},{\bar Q})$ that represents
a different spacetime.  The transformation is given by
\begin{eqnarray}
{\bar P}=\tau - P,
\label{GEP} \\
{{\bar Q}_{,\tau}}=- {e^{2(P-\tau)}}{Q_{,x}},
\label{GEQa} \\
{{\bar Q}_{,x}}=-{e^{2P}}{Q_{,\tau}}.
\label{GEQb}
\end{eqnarray}
This transformation is essentially the analog for Gowdy spacetimes of the  
Kramer-Neugebauer transformation\cite{kramer}
for stationary axisymmetric spacetimes.
The integrability conditions for equations (\ref{GEQa}-\ref{GEQb})
are satisfied as a result of the equations of motion 
(\ref{evolvePtau}-\ref{evolveQtau}).  Following\cite{alanandmarsha}
we will use the notation $({\bar P},{\bar Q})={\rm GE}(P,Q)$.  Here
the ``GE'' stands for ``Gowdy to Ernst.''  For our purposes what is
important is that if $(P,Q)$ has a false spike then 
${\rm GE}(P,Q)$ will have a true spike.  Thus from a solution 
$(P,Q)$ with no spikes at all we can generate a solution
${\rm GE}({\rm I}(P,Q))$ with true spikes.  The problem of proving
existence of solutions with spikes has thus been reduced to the problem
of proving existence of solutions with no spikes.  

How then does one prove existence of solutions with no spikes?
Here we simply sketch what is done and refer the reader to 
\cite{alanandk,alan} for details.  At first it might seem that what
is needed is a global existence theorem.  However, spikes are 
essentially asymptotic behavior as the singularity is approached.
Thus a local existence theorem will suffice provided that it is local
in a neighborhood of the singularity.  What is needed for such a 
theorem is to put the equations in Fuchsian form
\begin{equation}
t {{\partial {\vec u}}\over {\partial t}} + N(x){\vec u} = t
{\vec f}(t,x,{\vec u},{{\vec u}_{,x}})
\label{fuchs}
\end{equation}
for a system described by the vector valued function $\vec u$.  
One needs certain properties for the function $\vec f$ and matrix $N$.
Using these properties, one proves that there exists a solution in
a neighborhood of $t=0$ and with ${\vec u}=0$ at $t=0$.  To write
the Gowdy equations in Fuchsian form, one must first correctly guess
(based on equations (\ref{VTDPsoln}-\ref{VTDQsoln})) functions that
have the $t\to 0$ behavior of $P$ and $Q$.  One then writes $P$ and
$Q$ as these functions plus remainder terms.  One then rewrites the
Gowdy evolution equations (\ref{evolvePt}-\ref{evolveQt}) as
a Fuchsian system for the remainder terms.  In this way existence of
$P$ and $Q$ with the appropriate asymptotic behavior is proved.
The $P$ and $Q$ obtained in this way do not have spikes.  However, applying
the transformations I and GE to this $P$ and $Q$ one obtains a rigorous
proof of existence of Gowdy spacetimes with spikes.

\section{Conclusions}

When spikes were first found by Berger and Moncrief\cite{beverlyandvince}
they seemed quite mysterious.  Now, about a decade later, we have a
very good understanding of this phenomenon.  A variety of numerical
simulations have allowed us a close look at the process of spike
formation and at the properties of the spikes.  Analytical approximations
give us an understanding of how the spikes form as well as an
approximate description of their shape and time development.  
Mathematical results have shown that the false spikes are mere
coordinate artifacts while the true spikes are real geometric 
phenomena.  In addition it has been possible to prove rigorously
that solutions with spikes exist.

\ack

The work I have done on this subject has been done in collaboration
with Beverly Berger and Marsha Weaver.  I would like to thank them
as well as Vince Moncrief, Jim Isenberg and Alan Rendall
for numerous helpful discussions 
over the years.  This work was supported by NSF grant PHY-9988790
to Oakland University.

\end{document}